\documentclass{article}
\usepackage{arxiv}

\usepackage{graphicx}
\usepackage[toc,page]{appendix}

\usepackage{algorithm}
\usepackage[noend]{algpseudocode}
\usepackage{amsmath}
\usepackage{amsfonts}
\usepackage{apacite}
\usepackage{booktabs}
\usepackage{mathtools}

\newcommand{\beq}{\begin{equation}}
\newcommand{\eeq}{\end{equation}}
\newcommand{\bea}{\begin{eqnarray}}
\newcommand{\eea}{\end{eqnarray}}
\newcommand{\ba}{\begin{array}}
\newcommand{\ea}{\end{array}}
\newcommand{\bi}{\begin{itemize}}
\newcommand{\ei}{\end{itemize}}
\newcommand{\ben}{\begin{enumerate}}
\newcommand{\een}{\end{enumerate}}

\newcommand{\bfy}{\mbox{\boldmath{$y$}}}
\newcommand{\bfz}{\mbox{\boldmath{$z$}}}

\newcommand{\bfeta}{\mbox{\boldmath{$\eta$}}}

\newcommand{\bfepsilon}{\mbox{\boldmath{$\epsilon$}}}

\title{Sequential Bayesian Inference for Factor Analysis}

\author{
 Konstantinos Vamvourellis, Konstantinos Kalogeropoulos, Irini Moustaki\\
 Department of Statistics\\
  LSE\\
  \texttt{k.vamvourellis@lse.ac.uk}; \texttt{k.kalogeropoulos@lse.ac.uk}; \texttt{i.moustaki@lse.ac.uk}  \\
}

\mathtoolsset{showonlyrefs=true}
\begin{document}
\maketitle

\begin{abstract}

We develop an efficient Bayesian sequential inference framework for factor analysis models observed via various data types, such as continuous, binary and ordinal data. In the continuous data case, where it is possible to marginalise over the latent factors, the proposed methodology tailors the Iterated Batch Importance Sampling (IBIS) of \citeA{Chopin2002a} to handle such models and we incorporate Hamiltonian Markov Chain Monte Carlo. For binary and ordinal data, we develop an efficient IBIS scheme to handle the parameter and latent factors, combining with Laplace or Variational Bayes approximations. The methodology can be used in the context of sequential hypothesis testing via Bayes factors, which are known to have advantages over traditional null hypothesis testing. Moreover, the developed sequential framework offers multiple benefits even in non-sequential cases, by providing posterior distribution, model evidence and scoring rules (under the prequential framework) in one go, and by offering a more robust alternative computational scheme to Markov Chain Monte Carlo that can be useful in problematic target distributions.  
\end{abstract}

% keywords can be removed

\section{Introduction\label{sbfa_smc_sec:intro}}

Factor analysis is a statistical technique that uses a small number of latent factors to model the behaviour of a potentially larger number of observed variables. It can be used to model directional (regression coefficients) and non-directional (correlations) relationships amongst latent variables (structural model) which are identified by observed variables (measurement model). Factor analysis is part of Structural Equation Modelling (SEM) and Confirmatory Factor Analysis (CFA), where the focus is on verifying scientific hypotheses. Alternatively, Exploratory Factor Analysis (EFA) uses factor analysis as a dimension reduction technique or as a tool to uncover patterns in multivariate data. In both cases, Bayesian approaches have been developed and offer several benefits such as providing a natural framework for parsimonious model choice \cite{LW04,FL18,CFHP14}, performing well in small sample sizes \cite{DC15} and assessing model fit \cite{muthen2012}. 

Sequential Bayesian modelling has received attention in the setting of hypothesis testing where it has been proven superior to the traditional null hypothesis (NHST) paradigm. Specifically, Sequential Bayes Factors (SBF) do not suffer from bias associated to the stopping rule, the practice of stopping the processing of new data only when conclusive evidence is reached. Contrary to NHST theory, which requires a prespecified sampling plan, Bayes Factors allow for flexible sampling design and unlimited testing \shortcite{pramanik2021modified, schnuerch2020controlling}. One drawback of Bayes Factors is that they are not always easy to compute except from some simple cases of mean difference analysis \shortcite{schonbrodt2017sequential}. Several schemes have been proposed but they are only approximate and typically they are also non trivial to compute; see \citeA{VNM16} and the references therein for some examples in the context of factor analysis. The sequential scheme developed in this paper offers another alternative with the additional benefit that there is no need to refit the model from scratch when new data become available. 

The goal of a sequential scheme is to recursively explore the sequence of posterior distributions
\begin{equation}
\label{sbfa_smc_posteriors}
\pi_0(\theta) = p(\theta), \;\; \pi_i(\theta) = p(\theta | \bfy_{1:i}), \;\; i=1,\dots,n,
\end{equation}
where $\theta$ denotes all the unknown parameters in our model. The sequential inference paradigm approximates recursively these posterior distributions and the model evidence $p(\bfy_{1:i})$. For models where the likelihood $f(y|\theta)$ is available, the sequential scheme of  Iterative Batch Importance Sampling (IBIS) \cite{Chopin2002a} and its more general framework \cite{del2006sequential} provide the standard option. Factor models based on continuous and normally distributed data fall in this category and in this paper we tailor the IBIS algorithm for this case. For models including latent variables, such as factor models based on binary data, interest lies in the augmented joint posterior of parameters and latent variables 
\begin{equation}
\label{sbfa_smc_augmented_posteriors}
\pi_0(\theta, z_0) = p(\theta)  p(z), \;\; \pi_i(\theta, \bfz_{1:i}) = p(\theta, \bfz_{1:i} | \bfy_{1:i}), \;\; i=1,\dots,n,
\end{equation}
In such cases one option is provided by the $\text{SMC}^2$ scheme of \citeA{Chopin:2012gn} which focuses on the case where the latent variables satisfy the Markov property, e.g. Hidden Markov Models. In this paper we aim to construct an alternative scheme that takes advantage of the independence, rather than Markov dependence, between the latent variables that is typically assumed in factor analysis. 

Despite the fact that the developed computational scheme of the paper is sequential, it offers several benefits even in non-sequential contexts. First, it can provide the posterior joint distributions of the parameters and the model evidence in one go. Second, it allows the computation of scoring rules under the prequential framework; see for example \cite{DM14}. Third, it can provide a more robust alternative  computational scheme than MCMC schemes that can be helpful when the target distribution has problematic landscape, e.g. being multi-modal. 

We begin by laying out the framework for continuous data in Section \ref{sbfa_smc_sec:continuous_model}. We then proceed to explore the more challenging case of categorical data, given the presence of latent variables, in Section \ref{sbfa_smc_sec:binary_model}. In Section \ref{sbfa_smc_sec:applications} we demonstrate the framework with simulation experiments and a real data example. Finally Section \ref{sbfa_sec:discussion} concludes with some relevant discussion.

\section{Sequential Monte Carlo for Factor Analysis Based on Continuous Data}
\label{sbfa_smc_sec:continuous_model}

\subsection{Model and Priors}

We use a unified Bayesian framework that encompasses models for categorical and continuous observations. In this Section we focus on continuous data. Suppose there are $p$ observed variables (items) denoted by $\bfy=(y_1,\ldots,y_p)$ and that their associations are explained by $k$ continuous latent variables (factors) denoted by $\bfz = (z_1, \ldots, z_k)$. The classical linear factor analysis model ({\it{measurement model}}) is: %
\begin{equation}
\label{sbfa_smc_augmented}
\bfy_i = \alpha+\Lambda \bfz_i   + \bfepsilon_i, \;\; i=1,\ldots,n 
\end{equation}
where $\alpha$ is a $p\times 1$ vector of intercepts, $\Lambda$ is the $p\times k$ matrix of factor loadings and $n$ is the sample size. The vector of latent variables  $\bfz_i$ has usually a Normal distribution, $\bfz_i \sim N_k(0, \Phi)$. The $\bfepsilon_i$s are error terms assumed to be independent from each other and from the $\bfz_i$s. If they are assumed to be Normally distributed as $\bfepsilon \sim  N(0_p, \Psi), \;\;\Psi=\text{diag}(\psi_1^2,\dots,\psi_p^2)$, the latent variables $\bfz$ can be integrated out using standard properties of the Normal distribution, to obtain
\begin{equation}
\label{sbfa_smc_marginal}
\bfy_i\sim N(\alpha, \Lambda \Phi \Lambda^T+\Psi).
\end{equation}

The model defined in \eqref{sbfa_smc_augmented} and \eqref{sbfa_smc_marginal} applies to EFA, CFA or more generally SEM depending on how the parameters $\Lambda$ and $\Phi$ are formulated. EFA is the practice of factor analysis for the purpose of reducing the dimensionality of the observed outcomes. This is typically done by fitting a model where all the elements of $\Lambda$ are all free parameters, subject to some identifiability restrictions, while enforcing $\Phi = I_k$. In CFA and more broadly SEM, the focus is on establishing whether a hypothesised social, psychological, or other scientific theory, that determines which items load on each factor, is compatible with the data. Researchers express the theory by restricting certain elements of $\Lambda$ to zero (e.g. the so-called cross-loadings) to assign items to factors, and assessing the model's goodness of fit. The covariance matrix $\Phi$ is either unstructured (CFA) or defined by a parametric model that relates the latent variables with each other and optionally with the observed covariates (SEM). For both approaches it is common to set $\Psi$ to be a diagonal matrix, an assumption known as conditional independence of the variables given the factors.

Fitting the model requires resolving certain indeterminacies that arise. Since the scale of the latent factors is not identifiable there are generally two parametrisations to choose from: either the leading loadings in $\Lambda$ for each factor are fixed to a constant, usually $1$, or $\Phi$ is constrained to be a correlation matrix. In the second case, further care is needed as the sign of the factor loadings is not identifiable so the parameter space needs to be constrained to the positive or the negative side. Alternatively the parameter space could be unconstrained and instead post-sampling processing can be applied whereby the posterior samples of the loading columns corresponding to each factor should be multiplied by $-1$ if the relevant leading loading is negative, otherwise they are left as they are. This formulation may be viewed as a special case of the parameter expansion suggested in \citeA{GD09} for EFA. In the case of EFA further challenges have to be addressed due to the fact that the likelihood is specified in terms of $\Lambda \Lambda^T$ while the parameter of interest $\Lambda$ is free. Contrary to CFA, in the EFA setting there are no modelling constraints on the loading matrix, hence the likelihood is invariant under rotations of $\Lambda$. Enforcing $\Lambda$ to be lower triangular \cite<see e.g.>{GZ96} is one way to remove the rotational indeterminacy, but it introduces order dependence amongst the observed variables. The choice of the first $k$ variables, which is an important modelling decision \cite{CCLNWW08}, thus becomes inadvertently impactful. Alternative schemes have been proposed by \shortciteA{CFHP14,FL18,BD11}, which have the additional benefit of helping to identify the number of factors in a single MCMC run.

Regarding priors we start with the factor correlation matrix $\Phi$ which under the full covariance matrix parametrisation receives a prior with a low amount of information, e.g. the Inverse Wishart with the identity as the scale matrix and $p+4$ degrees of freedom or lower. Alternatively, if we use a correlation matrix $\Phi$, the LKJ prior is assigned, introduced in \shortciteA{LKJ09}, with a similar amount of low information, e.g. $\text{LKJ}(2)$. The free loadings in $\Lambda$ are assigned zero-centred Normal priors $N(0, \sigma^2)$. The prior variance is frequently set to be a large constant, which however can lead to issues related to Lindley's paradox \cite{L57}. An alternative choice that protects against such problems is the unit information priors \cite{KW96}, according to which variance is set to a small value that correspond to the amount of information from a single observation point. In this work, our items are in similar scales hence we fix the prior variance to $1$ in line with recommendations by \citeA{LW04} and \citeA{GD09}. Regarding the diagonal matrix $\Psi$, we assign independent Inverse Gamma priors on each $\psi_j^2$ 
$$ 
\psi_j^2 \sim \text{InvGamma}(c_0, (c_0-1)/(S_y^{-1})_{jj})
$$
where $S_y$ is the empirical covariance matrix and $c_0$ is a constant that the researcher can choose in order to limit the probability of running into Heywood issues as per recommendations in \shortciteA{FL18} and \shortciteA{CFHP14}. Finally, large variance Normal priors are assigned on the $\alpha$ parameters. In every analysis that follows we use the following wide prior Normal, $\alpha \sim N(0,10^2)$.

\subsection{Sequential Algorithm}

To target the recursive posteriors $\pi(\theta | \bfy_{1:i})$ we adopt the Iterated Batch Importance Sampling algorithm (IBIS), introduced by \citeA{Chopin2002a}. At a high level, IBIS works by propagating forward in time a set of parameter particles, each weighted by the likelihood function evaluated at its parameter values. For the standard IBIS approach we need to evaluate the likelihood function $f(y | \theta)$ which is available for the continuous case via formulation \eqref{sbfa_smc_marginal}. The ability to integrate out the latent variables, combined with the efficiency of the IBIS algorithm, achieves the goal we set in Section \ref{sbfa_smc_sec:intro}, i.e. to get an efficient process to draw samples from the sequence of posterior distributions $p(\theta | \bfy_{1:i}), \;\; i\geq 1$ and compute the model evidence. In what follows we describe the steps of the algorithm, while the full process is presented in Algorithm \ref{sbfa_smc_alg:ibis-cont}.
\begin{algorithm}
\caption{IBIS}\label{sbfa_smc_alg:ibis-cont}
Sample $\theta^{m}$, for $m=1,\dots,N_{\theta}$ from $\pi(\theta)$ and set $\omega^m=1$. All operations are assumed to be repeated for all $m\in 1: N_\theta$. \\
Then at time $i=1,\dots,n$, do: 
\begin{algorithmic}[1]
\State  Compute the incremental weights and their weighted average
$$
u_{i}(\theta^{m})=f(\bfy_{i}|\bfy_{i:i-1}, \theta^{m})=f(\bfy_{i}|\theta^m),\quad\quad
L_i=\frac{1}{\sum_{m=1}^{N_\theta}\omega^m}\times\sum_{m=1}^{N_\theta}\omega^m
 u_i(\theta^m),$$
\State  Update the importance weights
$$
\omega^{m}=\omega^{m}u_{i}(\theta^{m})
$$
\If{ESS$(\omega) < \gamma$} 
\Procedure{resample}{$\theta, \omega$}
\State \textbf{return} $\theta$
\EndProcedure
\Procedure{jitter}{$\theta^m, \bfy_{1:i}$}
    using an MCMC algorithm
\State \textbf{return} $\tilde\theta^m$
\EndProcedure
\State $(\theta^{m}, \omega^{m})= (\tilde{\theta}^{m}, 1)$
\EndIf
% \State \textbf{return} $\theta,\omega^m$\Comment{ready to move on to $t+1$}
\end{algorithmic}
\end{algorithm}
We begin by drawing $N_{\theta}$ samples of the parameter vector $\theta=(\alpha,\Lambda,\Phi,\Psi)$ from the prior distribution, called $\theta$ particles and denoted with $\{\theta_m\}_{m=1}^{N_{\theta}}$. We proceed in increments of time by considering one data point $\bfy_{i}$ at each time $i$. This is also known as data tempering and is adopted in this paper so that we can have assess the out of sample predictive performance of the model in question.

At each time $i$, or else data point, we compute the weights specified by the likelihood function $\omega_i^m=f(\bfy_{1:i} |\bfy_{1:i-1} \theta^m)$. Note that the latter simplifies to $f(\bfy_{1:i} | \theta^m)$ for the factor models defined in the previous subsection. Doing so, the weighted draws of the $\theta$ particles, $\{\theta^m,\omega_{i}^m\}_{m=1}^{N_{\theta}}$ at time $i$ can be used to evaluate summaries of the posterior $\pi(\theta | \bfy_{1:i}) $. More specifically, expectations with respect to that posterior, $E[g(\theta)|\bfy_{1:i}]$, can be computed using the estimator 
\begin{equation}
\frac{\sum_{m}[\omega_m g(\theta^m)]}{\sum_{m}\omega_m}\rightarrow E[g(\theta)|\bfy_{1:i}].
\end{equation}
 \cite{Chopin2004} shows consistency and asymptotic normality of this estimator as $N_{\theta}\rightarrow \infty$ for all appropriately integrable $g(\cdot)$. The same holds for expectations with respect to the posterior predictive distributions such as
$f(\bfy_{i+1}|\bfy_{1:i})$. Since
 $$
 f(\bfy_{i+1}|\bfy_{1:i})\propto  f(\bfy_{i+1}|\bfy_{1:i},\theta)\pi(\theta|\bfy_{1:i}),
 $$
the weighted $\theta$ particles, from $\pi(\theta|\bfy_{1:i})$, can be transformed into weighted $y_{i+1}$ particles from $f(y_{i+1}|\bfy_{1:i})$ by simply drawing $y_{i+1}^m$ from $f(y_{i+i}|\bfy_{1:i},\theta^m)$ which, as mentioned earlier, equals to $f(y_{i+1}|\theta^m)$ in our case. Moreover, a very useful by-product of the IBIS algorithm is the ability to compute the model evidence $f(\bfy_{1:i})$, to calculate Bayes factors. Computing the following quantity in step 1 in Algorithm \ref{sbfa_smc_alg:ibis-cont} yields a consistent and asymptotically normal estimator of $f(\bfy_{i}|\bfy_{1:i-1})$
\begin{equation}
\frac{1}{\sum_{m=1}^{N_{\theta}}\omega^{m}}\sum_{m=1}^{N_{\theta}}\omega^mu_{i}(\theta^m)\rightarrow f(\bfy_{i}|\bfy_{i:i-1}).
\end{equation}
In other words the output of the IBIS output allows the calculation of all the summaries often obtained from the MCMC outputs, such as the posterior mean, mode, or median, 95\% credible intervals, samples from the predictive distribution, but for all the posteriors $\pi(\theta|\bfy_{1:i})$, $i=1,\dots,n$. Moreover it provides estimates of the model evidence for all $i$. 

Note that if we were to only propagate the particles according to steps 1 and 2 of the (IBIS) Algorithm  \ref{sbfa_smc_alg:ibis-cont}, the weights of the particles will eventually deteriorate with very few or even one of them dominating the others, which will lead to inaccurate estimates of the posterior summaries. One index that measures the quality of the weighted $\theta$ particles is the effective
sample size (ESS)
\begin{equation}
  \label{sbfa_smc_eq:ess}
  \mathrm{ESS}(\omega)=\frac{\left(\sum_{m=1}^{N_\theta}\omega^{m}\right)^{2}}{\sum_{m=1}^{N_\theta
}\left(\omega^{m}\right)^{2}}.
\end{equation} 
The protocol of the IBIS algorithm requires to monitor a degeneracy criterion, which is typically to check if the ESS is less than a prespecified threshold $\gamma$, the violation of which triggers a two-step procedure to improve the quality of the $\theta$ particles. The first step of this procedure is to resample the $\theta$ particles with replacement, e.g. via the multinomial distribution with the normalised weights as probabilities. At that point we reset all weights to $1$ but we end up having multiple copies of the $\theta$ particles with high weights, whereas some $\theta$ particles with low weights are removed. The purpose of this step is to drop $\theta$ particles of low weights and focus on the ones with high weight. This can be particularly helpful in the presence of local modes, since the $\theta$ particles that can potentially get trapped there will eventually be removed if the density at those modes is low. The second step of this procedure, called jittering, is to apply a MCMC algorithm with initial value at each $\theta^m$ particle to sample from the posterior given data up to that point. The MCMC algorithm is run for a few iterations and the last value of the MCMC chain, denoted by $\tilde\theta^m$ becomes the new value $\theta^m$. The purpose of jittering is to avoid having exact multiple copies in the set of $\theta$ particles and the use of MCMC ensure that the desirable asymptotic properties of the IBIS output are not violated; see \cite{Chopin2002a,Chopin2004,del2006sequential} for details on the relevant theory.

Hence, in order to fully define the IBIS algorithm, it necessary to provide a MCMC algorithm to sample from the posteriors $\pi(\theta|\bfy_{1:i})$ for all $i$. Note that the standard Gibbs sampler of \cite{GZ96} is not immediately suitable for this purpose as it returns samples from $\pi(\theta,\bfz|\bfy_{1:i})$. We therefore proceed with Hamiltonian MCMC targeting the posterior based on the likelihood in \eqref{sbfa_smc_marginal} where the latent factors have been marginalised out. The Hamiltonian MCMC algorithm can be applied using the standard publicly available platform Stan \shortcite{stan}. The code used for this paper, which is provided in the accompanying repository\footnote{https://github.com/bayesways/smc2}, combines the IBIS algorithm with the use of PyStan, the Python interface of the Stan language. The fact that the MCMC and IBIS target $\pi(\theta|y)$, as opposed to  $\pi(\theta,\bfz|\bfy_{1:i})$, does not imply that it is no longer possible to explore the posterior of the latent factors. Note that under the model of \eqref{sbfa_smc_augmented} and for all $i\leq j \leq n$
$$
\pi(\bfz_i|\bfy_{1:j})\propto \pi(\bfz_i| \theta,\bfy_{1:j})\pi(\theta|\bfy_{1:j})=\pi(\bfz_i| \theta,\bfy_{i})\pi(\theta|\bfy_{1:j}),
$$
which for the case of $\Phi = I_k$ is $(\bfz_i| \theta,\bfy_{i}) \sim N \big( (I_k + \Lambda^T \Psi^{-1} \Lambda)^{-1} \Lambda^T \Psi^{-1} \bfy_{i} \;,\;  (I_k + \Lambda^T \Psi^{-1} \Lambda)^{-1} \big)
$, see e.g. \shortciteA{GZ96, LW04}. A similar expression is available for the general case. Hence the IBIS output can be used to transform the $\theta$ particles into ${\bfz_i}$ particles by simply drawing from the full conditional above for each $\theta^m$ given a data point $\bfy_{i}$.

From a computational point of view, the most expensive step of the IBIS algorithm is the jittering step that requires to run an MCMC routine for a few iterations per $\theta$ particle. This is roughly equivalent to running several MCMC algorithms based on the smaller dataset $\bfy_{1:i}$ for some $i$. Nevertheless, note that jittering is more likely to occur when, in the transition between $\pi(\theta|\bfy_{1:i-1})$ and $\pi(\theta|\bfy_{1:i})$, these two posteriors are substantially different. As a consequence, most of jittering steps tend to take place for small $i$s, where the learning curve is steeper, and become less frequent as $i$ increases. This suggests that the computational cost of the IBIS algorithm is typically larger than running a single MCMC algorithm on the full data $y_{1:n}$ but usually not by much. The difference can often be eliminated by using parallel computing; more specifically running the MCMC chains of the jittering step in parallel for each $\theta$ particle. For more coding details see Appendix \ref{sbfa_app:code} and the accompanying repository.

\section{Sequential Monte Carlo methods for Binary data\label{sbfa_smc_sec:binary_model}}

\subsection{Model, Priors and MCMC Scheme}

Binary and ordinal type data can be accommodated by extending the model framework used for continuous data and viewing the categorical responses as manifestations of underlying (latent) continuous variables denoted by ${\bfy^*} = (y^*_{1}, \ldots, y^*_{p})$. When continuous variables are analysed, $y_j=y^*_j$, ($j=1,\ldots,p$). The classical linear factor analysis model in the general form then becomes: %
\begin{equation}
\label{sbfa_smc_bin_augmented}
\bfy^*_i = \alpha+\Lambda \bfz_i   + \bfepsilon_i, \;\; i=1,\ldots,n 
\end{equation}

For binary data, the connection between the observed binary variable $y_j$ and the underlying variable $y^*_j$ is $y_{j}=\mathcal{I}(y_{j}^*>0)$. Specific choices for the distribution of the error term $\bfepsilon$ lead to the following well known models:
\begin{equation}
\label{sbfa_smc_bin_augmented2}
\bfepsilon\sim\begin{cases}  N(0_p, \Psi), \;\;\Psi=\text{diag}(\psi_1^2,\dots,\psi_p^2), & \text{if }\bfy_i\text{ is continuous}\\
N(0_p, \Psi), \;\;\Psi= I_p, &\text{if }\bfy_i\text{ is binary and the probit model is adopted}\\
 \prod_{j=1}^J\text{Logistic}(0,\pi^2/3),  & \text{if }\bfy_i\text{ is binary and the logit model is adopted},
\end{cases}
\end{equation}
where $0_p$ is a $p$-dimensional vector of zeros and $I_p$ denotes the identity matrix of dimension $p$. The marginal distribution of the underlying variable becomes:
\begin{equation}
\label{sbfa_smc_bin_marginal}
\bfy^*_i\sim N(\alpha, \Lambda \Phi \Lambda^T+\Psi).
\end{equation}

The expression above is equivalent to the following
\begin{equation}
\label{sbfa_binaug}
 \begin{cases}
{\bf y}_{i}\sim \prod_{j=1}^p\text{Bernoulli}\big(\pi_{ij}(\eta_{ij})\big)\\
\pi_{ij}(\eta_{ij})=\sigma(\eta_{ij}) \;\;\text{or}\;\; \pi_{ij}(\eta_{ij})=\Phi(\eta_{ij}),\;\;\eta_{ij}=[\bfeta_i]_j\\
{\bfeta}_i \coloneqq \alpha+\Lambda {\bf z}_i,\\
{\bf z}_i \sim N(0,\Phi)
\end{cases}
\end{equation}
where $\sigma(\cdot)$ denotes the sigmoid function and leads to the logit model, whereas $\Phi(\cdot)$ denotes the cumulative density function of the standard Normal distribution and leads to the probit model.

Priors are set similarly to the continuous data case described in Section \ref{sbfa_smc_sec:continuous_model}. Parameters $\alpha$ and $\Phi$ are treated identically. However, the unit information prior for the loading parameters $\Lambda$ could be different.  In the case of the 2PL IRT model this translates to a $N(0,4)$ prior \cite{VNM14}, which is the one we adopt for most models as well. Finally, in the binary data case there is no longer a matrix $\Psi$ to consider under formulation \eqref{sbfa_binaug}.

In order to sample from the posterior given all the available data, the Gibbs sampler is only available for the case of the probit link \cite{CG98}, whereas a Metropolis withing Gibbs algorithm has also been used, see \citeA{VNM16} and the references therein. Aiming for an efficient and general option, we again resort to Stan, which can in principle be used for all such models. Stan was also used in \shortciteA{vamvourellis2021generalised} where it performed reasonably well. As described in the previous Section, an efficient MCMC scheme is an essential ingredient for sequential schemes, the topic we turn to next.

\subsection{Sequential Scheme}\label{sbfa_smc_subsec:seqschemes}

The IBIS approach taken in the continuous data is not directly relevant for the case of binary data because we cannot integrate the latent factor to obtain a likelihood of the form $f(y | \theta)$ for categorical data, except for the case of the probit link. Instead, we can easily evaluate the augmented likelihood $f(y\;|\; z, \theta)$ according to \eqref{sbfa_smc_bin_augmented2} where the underlying variable is given by \eqref{sbfa_binaug} depending on the model chosen. There are two potential routes in order to construct a sequential Monte Carlo scheme that includes both parameters $\theta$ and latent variables $\bfz_{1:i}$. The first is to consider an IBIS algorithm on the higher dimensional parameter vector, that includes both parameter and latent variables, and explore the properties of factor analysis models to mitigate potential issues caused by the increased dimension. An example application of the IBIS algorithm in high dimensions is provided by \cite{KBJ14}. The second route is to pursue the development of a scheme in the spirit of the $\text{SMC}^2$ of \cite{Chopin:2012gn}. SMC$^2$ focuses mostly on the case of Markov dependent latent variables and combines the IBIS algorithm with the pseudo marginal framework of \citeA{andrieu2009pseudo} and, more specifically, the particle MCMC algorithm \cite{andrieu2010particle}. In this paper we proceed along the lines of the former route, by constructing an efficient IBIS algorithm on the augmented parameter vector $\Theta=(\theta,\bfz_{1:n})$, enhanced with importance sampling targeting the posterior of the latent factor $\bfz_{1:n}$ based on the Laplace or Variational Bayes approximations \cite{blei2017variational}. Note that the developed approach requires a single $z$ particle for each $\theta$ particle, as opposed to $N_z$ particles in the SMC$^2$ framework. 

Getting to the specifics of the IBIS algorithm, we take Algorithm \ref{sbfa_smc_alg:ibis-cont}, formulated on $\Theta$ rather than $\theta$, as starting point and present our modifications that lead to the Algorithm \ref{sbfa_smc_alg:ibis-lvm}, which we recommend in this paper for the case of binary data.  The initialisation in the case of Algorithm \ref{sbfa_smc_alg:ibis-cont}, requires to draw samples from the priors of $\theta$ and $\bfz_i$ for $i=1,\dots,n.$ Note however, that at time $i$ of the IBIS the latent factors $\bfz_j$ for $i<j<n$ do not contribute at all to the algorithm; neither in the incremental weights nor in the likelihood for the MCMC in the jittering step. They can therefore be omitted and drawn retrospectively when the algorithm reaches $i=j$. 

One of the key features regarding the efficiency of the IBIS algorithm is the number of times the degeneracy criterion is triggered. Under the data tempering schedule the ESS is being reduced at time $i$ due to the differences between the posteriors based on $\bfy_{1:i-1}$ and $\bfy_{1:i}$. When it comes to $\theta$ such differences typically become smaller as $i$ increases. This is not the case, however, for latent factors $\bfz_i$s. As noted earlier, for factor analysis models, $\pi(\bfz_i|y_{1:n}, \theta)=\pi(\bfz_i|\bfy_{i}, \theta)$, which implies that given $\theta$ the learning regarding $\bfz_i$ takes place only at time $i$. Hence, if the posteriors $\pi(\bfz_i|\bfy_{i}, \theta)$s tend to be substantially different than the priors, the degeneracy criterion of Algorithm \ref{sbfa_smc_alg:ibis-cont} may end up being triggered at all times, thus leading to a very inefficient computational scheme. The problem can potentially be addressed by replacing data tempering with another schedule, but this will no longer enable desirable features such as sequential testing and evaluation of scoring rules. We therefore proceed by retaining the data tempering schedule and resort to importance sampling to address this issues. More specifically, rather than drawing each $\bfz_i$ from its prior, which leads to incremental weights $f(\bfy_{i}|\bfy_{1:i-1},\theta,\bfz_i)=f(\bfy_{i}|\theta,\bfz_i)$, we draw them from a proposal distribution $q(.)$ and compute the weights according to 
\begin{equation}
\label{sbfa_smc_ibis-weight}
u_i(\bfz^m_{1:i},\theta^m) = \frac{f(\bfy_{i}| \theta^m , \bfz_{1:i}^m) \pi(z)}{q(\bfz^m_i | \theta^m, \bfy_{1:i} )}
\end{equation}
We seek a proposal $q(\cdot)$ that resembles the posterior $\pi(\bfz_i | \theta^m, \bfy_{1:i})=\pi(\bfz_i | \theta^m, \bfy_{i})$. The density of this posterior is not available in closed form but it is typically low dimensional and its likelihood function consists of a single data point. Hence, it is not hard neither computationally expensive to obtain approximations of it, such as the  Laplace method which is easy to program in the presence of the relevant derivatives; see  Appendix \ref{sbfa_smc_appendix:laplace}. Another option is provide by Variational Bayes (VB) \shortcite{kingma2013auto,kucukelbir2017automatic} to derive a distribution that approximates the posterior $\pi(\bfz_i|\bfy_{i}, \theta)$. The VB method has the advantage of not requiring the model specific expressions of the derivatives and can be automated.

We now summarise and present the full process  in Algorithm \ref{sbfa_smc_alg:ibis-lvm}. The main differences of this IBIS algorithm when contrasted with the Algorithm \ref{sbfa_smc_alg:ibis-cont} is the augmentation of the particles with latent variables, thus having $\{\Theta^m\}_{m=1}^{N_{\Theta}}$ particles, the usage of the augmented likelihood in \eqref{sbfa_smc_augmented} instead of the marginal in \eqref{sbfa_smc_marginal} and the incremental addition of the latent factor particles.  

\begin{algorithm}
\caption{IBIS-LVM}\label{sbfa_smc_alg:ibis-lvm}
Sample $\theta^{m}$ from $\pi(\theta)$ and set $\omega^m=1$ for $m\in 1: N_{\Theta}$. All operations are assumed to be repeated for all $m\in 1: N_{\Theta}$. \\
Then at time $i=1,\dots,n$ do: 

\begin{algorithmic}[1]
\State Sample $\bfz_{i}^m \sim q(\bfz_{i}|\bfy_{i}, \theta^m)$
\State Append $\bfz_{i}^m$ to $\bfz_{1:i-1}^m$ to maintain the matrix of latent variables (if $t=1$ initialise matrix $\bfz_{1:1}^m = \bfz_1^m$ )
\State  Compute the incremental weights and their weighted average
$$
u_{i}(\theta^{m}, \bfz_{i}^m)= \frac{f(\bfy_{i}|\theta^{m},\bfz_{i}^m)\pi(\bfz_{i}^m|\theta^{m})}{q(\bfz_{i}^m|\bfy_{i},\theta^{m})} ,\quad
L_i=\frac{1}{\sum_{m=1}^{N_\theta}\omega^m}\times\sum_{m=1}^{N_\theta}\omega^m
 u_i(\theta^m, \bfz_{i}^m),$$
\State  Update the importance weights
$$
\omega^{m}= \omega^{m}u_{i}(\theta^{m}, \bfz_{i}^m)
$$
\If{ESS$(\omega) < \gamma$} 
\Procedure{resample}{$\theta^m, \bfz^m_{1:i}, \omega$}
\State \textbf{return} $\theta^m$, $\bfz^m_{1:i}$
\EndProcedure
\Procedure{jitter}{$\theta^m, \bfz_{1:i}^m, \bfy_{1:i}$}
    using an MCMC algorithm
\State \textbf{return} $\tilde\theta^m, \tilde \bfz_{1:i}^m$
\EndProcedure
\State $(\theta^{m}, \bfz_{1:i}^m,  \omega^{m})= (\tilde{\theta}^{m},  \tilde \bfz_{1:i}^m , 1)$
\EndIf
% \State \textbf{return} $\theta,\omega^m$\Comment{ready to move on to $t+1$}
\end{algorithmic}
\end{algorithm}

\section{Applications}\label{sbfa_smc_sec:applications}

We perform sequential parameter inference and model choice in simulated and real datasets considering cases of continuous and categorical data. We start with simulations of continuous data in Section \ref{sbfa_smc_subsec:contsim} to show that the final posterior distribution are correctly recovered. We then extend those results to the case of binary data in Section \ref{sbfa_smc_subsec:binsim} where we also compare the suggested methodologies from Section \ref{sbfa_smc_sec:binary_model}. Section \ref{sbfa_smc_subsec:model_choice} focuses on Bayes Factors for sequential model choice in different scenarios including  choosing the number of factors. Finally, Section \ref{sbfa_smc_subsec:big5} contains a real data example from the British Household Panel Survey.

\subsection{Continuous Data Simulations}\label{sbfa_smc_subsec:contsim}

We generated data $\bfy$ from the model of equation \eqref{sbfa_smc_augmented} with two factors according to the following process, denoted as `Continuous Scenario 1': the sample size $n$ is $200$, and $\alpha$ is a vector of zeros, the factor scores $\bfz_i$s were generated from the $N(0, \Phi)$, whereas the error terms were distributed according to the  $N(0, \Psi)$ with the associated parameters $\Lambda$, $\Psi$ and $\Phi$ being as follows
\begin{equation}\label{sbfa_smc_table:cont_sim}
\begin{split}
    \Psi &= \text{diag}(0.35, 0.58, 0.58, 0.35, 0.58, 0.58) \\
    \Phi &= 
    \begin{pmatrix}
    0.65 & 0.13 \\
    0.13 & 0.65
    \end{pmatrix} \\
    \Lambda &= 
    \begin{pmatrix}
    1.0 & 0 \\
    0.8 & 0 \\
    0.8 & 0 \\
    0 & 1 \\
    0 & 0.8 \\
    0 & 0.8 \\
    \end{pmatrix} 
\end{split}
\end{equation}
A CFA model was fit to the simulated data where the loading matrix structure mirrors the loading matrix used in the data generation process, in other words the locations of zeros were assumed to be known but the remaining loading were considered as unknown parameters. Moreover, the residual covariance matrix was restricted to be diagonal.  

\begin{figure}
\centering
\includegraphics[width=0.8\textwidth]{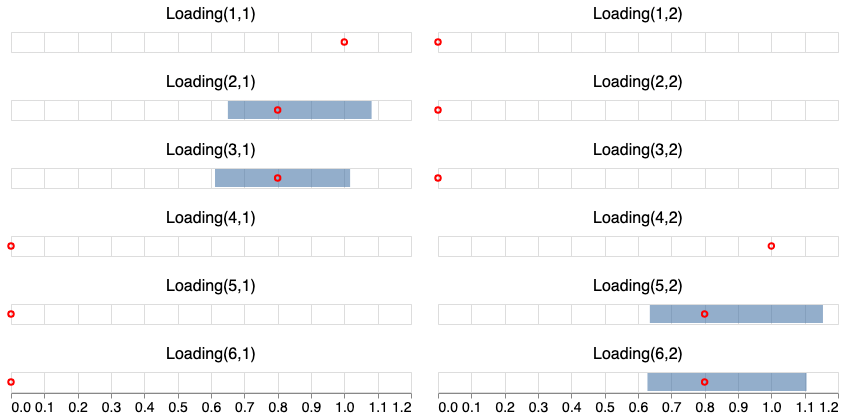}
\caption{Real data values marked in red dots overlaid with the  95\% credible intervals for $\Lambda$, the loading matrix parameters in Continuous Scenario 1.}
\label{sbfa_smc_fig:ezsim_credint1}
\end{figure}

\begin{figure}
\centering
\includegraphics[width=0.4\textwidth]{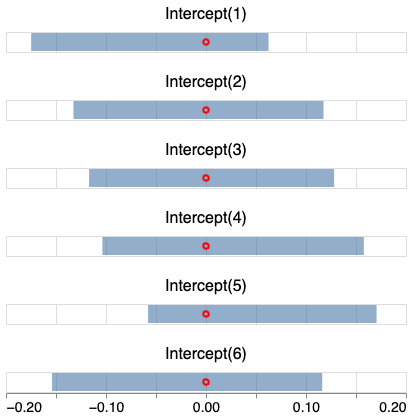}
\caption{Real data values marked in red dots overlaid with the  95\% credible intervals for $\alpha$, the intercept  parameters in Continuous Scenario 1.}
\label{sbfa_smc_fig:ezsim_credint2}
\end{figure}

\begin{figure}
\centering
\includegraphics[width=0.8\textwidth]{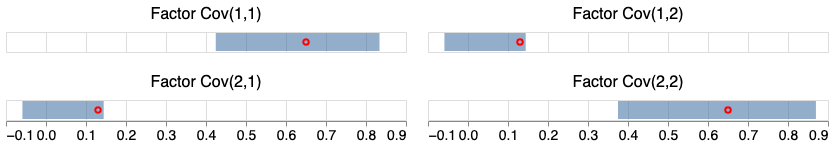}
\caption{Real data values marked in red dots overlaid with the  95\% credible intervals for $\Phi$, the factor covariance parameters in Continuous Scenario 1.}
\label{sbfa_smc_fig:ezsim_credint3}
\end{figure}

\begin{figure}
\centering
\includegraphics[width=0.4\textwidth]{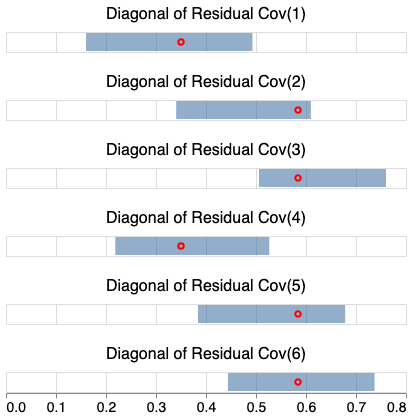}
\caption{Real data values marked in red dots overlaid with the  95\% credible intervals for $\text{diag}(\Theta)$ the diagonal elements of the residual covariance parameters in Continuous Scenario 1.}
\label{sbfa_smc_fig:ezsim_credint4}
\end{figure}

\begin{figure}
\centering
\includegraphics[width=0.8\textwidth]{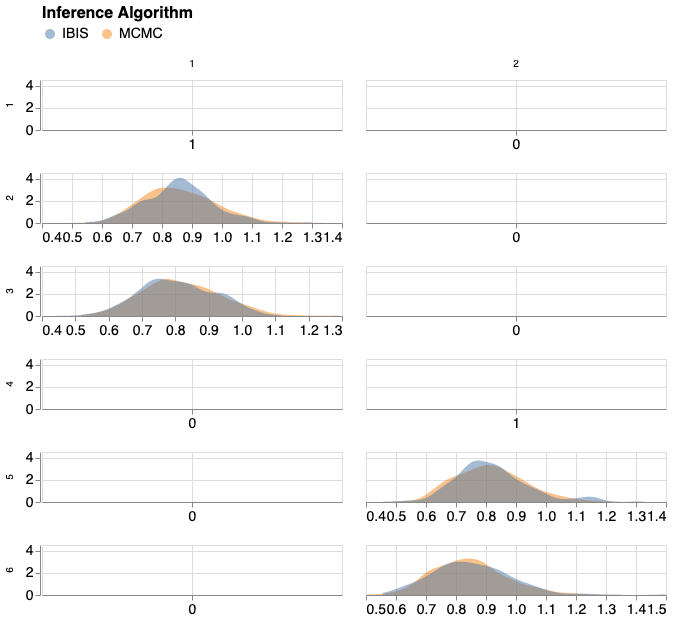}
\caption{Posterior Draws for the Loading matrix $\Lambda$ in Continuous Scenario 1, for both IBIS and batch MCMC methodologies.}
\label{sbfa_smc_fig:ezsim_loadings}
\end{figure}

\begin{figure}
\centering
\includegraphics[width=0.4\textwidth]{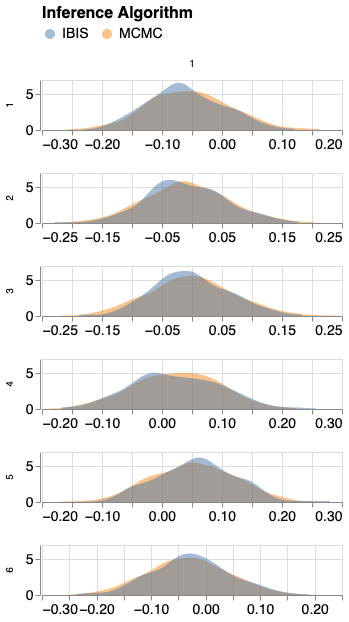}
\caption{Posterior Draws for the Intercept parameters $\alpha$ in Continuous Scenario 1, for both IBIS and batch MCMC methodologies.}
\label{sbfa_smc_fig:ezsim_intercept}
\end{figure}

\begin{figure}
\centering
\includegraphics[width=0.8\textwidth]{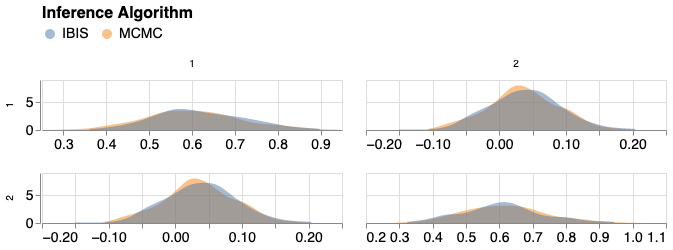}
\caption{Posterior Draws for the Factor Covariance matrix $\Phi$ in Continuous Scenario 1, for both IBIS and batch MCMC methodologies.}
\label{sbfa_smc_fig:ezsim_factorcov}
\end{figure}

\begin{figure}
\centering
\includegraphics[width=0.4\textwidth]{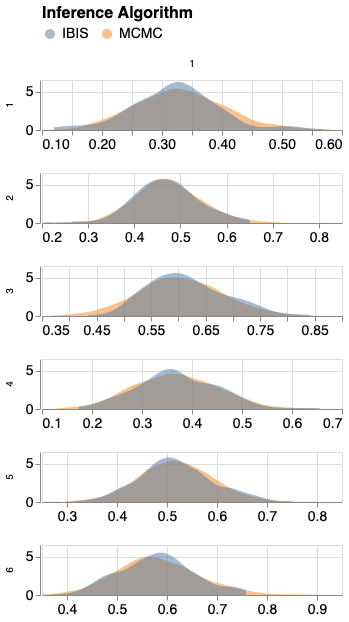}
\caption{Posterior Draws for the diagonal of the covariance matrix of the residual Errors $\text{diag}(\Theta)$ in Continuous Scenario 1, for both IBIS and batch MCMC methodologies.}
\label{sbfa_smc_fig:ezsim_diagtheta}
\end{figure}

We performed inference using the IBIS algorithm as presented in Section \ref{sbfa_smc_sec:continuous_model} and recovered posterior samples for all parameters at each step $i=1,\ldots,n$. In Figures  \ref{sbfa_smc_fig:ezsim_credint1}, \ref{sbfa_smc_fig:ezsim_credint2}, \ref{sbfa_smc_fig:ezsim_credint3}, and \ref{sbfa_smc_fig:ezsim_credint4}, we plot the 95\% credible intervals after processing the last data point ($i = 200$) and verify that they include the correct values (marked with red dots). Note that for identifiability reasons the leading loading parameters were fixed to $1$, hence the free loading parameters to be estimated in this model are the $4$ elements in the positions containing  the values $0.8$ in $\Lambda$ matrix in \eqref{sbfa_smc_table:cont_sim}. We also fit the same model using a batch MCMC algorithm to the same exact data set and compared the posterior draws. More specifically, we compared the posterior draws of the sequential IBIS methodology with the posterior draws from an MCMC batch algorithm that was fit to the full dataset all at once. We confirmed that the posterior draws are essentially identical, which demonstrates the correctness of the sequential methodology. In Figures \ref{sbfa_smc_fig:ezsim_loadings}, \ref{sbfa_smc_fig:ezsim_intercept}, \ref{sbfa_smc_fig:ezsim_factorcov}, and \ref{sbfa_smc_fig:ezsim_diagtheta}, we present the density plots of the loadings parameters overlaying the density charts for both algorithms. We see that the charts coincide almost completely, which shows that at the end of the process the sequential algorithm  produces the same posterior density plots as the standard MCMC batch algorithm.

\subsection{Parameter Estimation for Binary Data}\label{sbfa_smc_subsec:binsim}

We now proceed to demonstrate the sequential inference algorithm for a case of simulated binary data. We run three inferential processes on the same data set, each time using one of the three methodologies presented in Section \ref{sbfa_smc_subsec:seqschemes}, sampling the latent variables from their prior, the Laplace approximation of their posterior, and the Automatic Variational Bayes (ADVI)  approximation \cite{kucukelbir2017automatic} of their posterior. The data set was generated from a single factor model based on formulation \eqref{sbfa_binaug} adopting the logit link and using the following parameter values: $n$ = 100, $\alpha =(-0.53,  0.35, -1.4 , -1.4 , -0.96, -2.33)$, the loading matrix $\Lambda = (1,1,1,1,1,1)$, and the factor scores were generated as $z \sim N(0, 1)$. We call this data generation process `Binary Scenario 1'.

\begin{figure}
\centering
\includegraphics[width=0.4\textwidth]{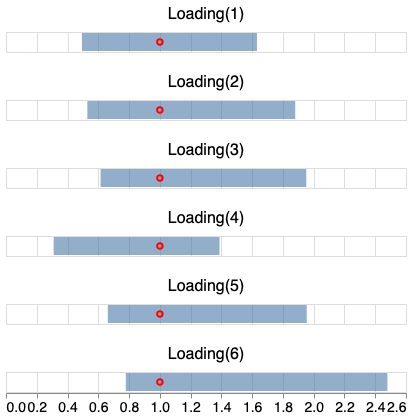}
\caption{Real data values marked in red dots overlaid with the  95\% credible intervals for the loading matrix parameters $\Lambda$ in the Binary Scenario 1 data simulation.}
\label{sbfa_smc_fig:binsim_credint1}
\end{figure}

\begin{figure}
\centering
\includegraphics[width=0.4\textwidth]{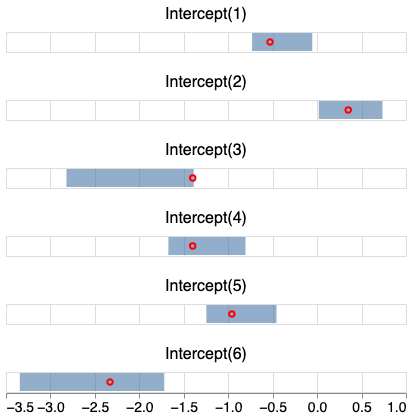}
\caption{Real data values marked in red dots overlaid with the  95\% credible intervals for the intercept parameters $\alpha$ in the Binary Scenario 1 data simulation.}
\label{sbfa_smc_fig:binsim_credint2}
\end{figure}

\begin{figure}
\centering
\includegraphics[width=0.5\textwidth]{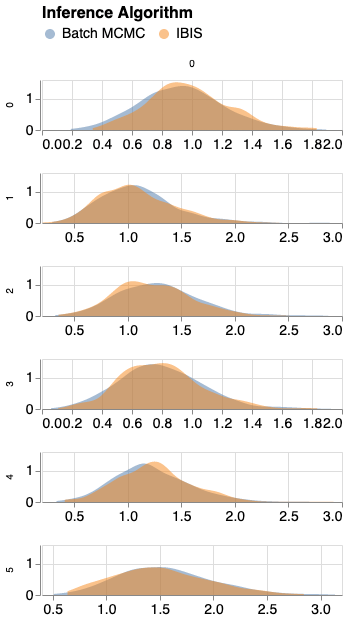}
\caption{Posterior Draws for the Loading matrix $\Lambda$ in for EZ model in the binary data simulation, for both IBIS and batch MCMC methodologies.}
\label{sbfa_smc_fig:binsim_loadings}
\end{figure}

\begin{figure}
\centering
\includegraphics[width=0.5\textwidth]{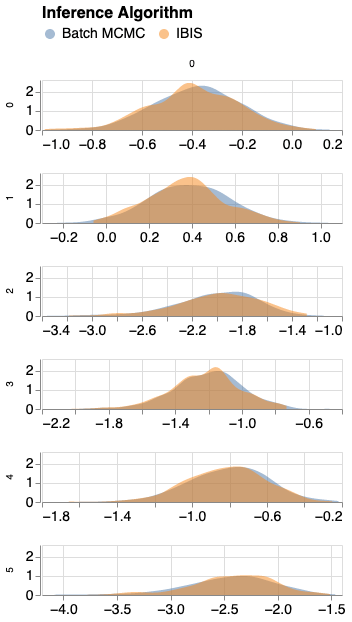}
\caption{Posterior Draws for the intercept parameters $\alpha$ of EZ model in the binary data simulation, for both IBIS and batch MCMC methodologies.}
\label{sbfa_smc_fig:binsim_intercepts}
\end{figure}

The simulations demonstrate that the naive approach of simulating from the prior, denoted by `PRIOR' is inefficient. We contrast PRIOR to the two alternative methods we propose in this paper, namely the Laplace approximation method (LAPLACE) and the Variational Bayes method (VB). The PRIOR method is relatively easy to implement because the latent variables are drawn from the prior. The drawback of this method is that because the latent particles come from the prior, very few of them achieve high likelihood values. As a result, we need to refresh the particles very often by running a full MCMC chain for each particle, which is typically the most costly step in a sequential algorithm. In the 100 data point data set we used, the ESS criterion was triggered  57 times under the PRIOR scheme. The problem is particularly acute because the resampling and jittering rate does not necessarily drop as the data index $i$ increases, contrary to the other methods we examined. More specifically, the degeneracy criterion was triggered 30 times in the first 50 data points and 27 in the last 50 data points; in other words the rate fell by only 10\% in the second half. The LAPLACE methodology requires extra work to derive the derivatives but is far more efficient in terms of resampling rate. More specifically, the ESS criterion was triggered only 27 times in the 100 data points. More importantly, the resampling and jittering rate in the second half of the data set was (9/50) whereas in the first half it was (18/50), thus resulting in 50\% reduction. Regarding the additional run time cost of computing the Laplace approximation, we found it to be minor in our simulations. Informally, we note that in the experiments we ran the run time of LAPLACE method was only 10-20\% higher compared to the PRIOR method, in the order of added minutes. A potential drawback of this scheme is that the Laplace approximation requires the analytical derivatives, which may not be simple to derive. A more automatic approach is offered by the VB method which does not require the user to manually derive the derivative formulas needed for the Laplace approximation. The efficiency of VB in our simulations was exactly the same as that of LAPLACE, with the ESS criterion being triggered at 18/50 times in the first half and 9/50 times in the second half of the data points. One possible drawback of the VBA is the extra run time needed to compile and run the VBA step for each draw. While the exact run time is highly dependent on the software and hardware used, to reduce run time we advise using a language that can run compiled models in order to amortise the compile time of the model across the rest of the resampling steps. In Figures \ref{sbfa_smc_fig:binsim_credint1} and \ref{sbfa_smc_fig:binsim_credint2}   we plot the 95\% credible intervals for the loading parameters along with the real data values and verify that the credible intervals cover the true values used to generate the data. Additionally, just as we did with the continuous data simulations, we fit both the batch MCMC and the sequential algorithm
to the same dataset and compare the posterior draws. We verify that the draws of the sequential framework at the last step coincide with those from the batch MCMC algorithm fit to the entire data set. We present the comparison in Figures \ref{sbfa_smc_fig:binsim_loadings} and \ref{sbfa_smc_fig:binsim_intercepts}. For simplicity we show the output of the LAPLACE method only since the rest of the methods produce similar results. 

\subsection{Sequential Model Choice}\label{sbfa_smc_subsec:model_choice}

\subsubsection{Overview}

In this Section, and also the next one, we demonstrate the use of Sequential Bayes Factors in order to facilitate model choice. We include special considerations for choosing the right number of factors. After we have briefly introduced the concept, we lay out the range of candidate models. We then proceed to run simulations using synthetic continuous data, which allows us to check our approach since we know the best model in advance. 

The fully Bayesian approach to model choice is based on computing the model evidence and the probability of the observed data under the model of choice. The evidence of model $\mathcal{M}$, also called marginal likelihood, is defined as follows

\begin{equation}
\pi(y | \mathcal{M}) = \int f(y|\theta, \mathcal{M}) \pi(\theta, \mathcal{M}) d\theta
\end{equation}

where $f(y|\theta, \mathcal{M})$ is the likelihood function under the model $\mathcal{M}$, and $\pi(\theta, \mathcal{M})$ is the prior distribution under model $\mathcal{M}$. When the model is implied, we omit it from the formulas for ease of notation. In the Bayesian framework, to compare two models $\mathcal{M}_1$ and $\mathcal{M}_2$ is to compare their evidence. The ratio of their evidence is the Bayes Factor which we will abbreviate as 
$$
BF(\mathcal{M}_1/\mathcal{M}_2) \coloneqq \frac{\pi(y | \mathcal{M}_1)}{\pi(y | \mathcal{M}_2)}
$$

The Bayes Factor of two models provides a numerical comparison as for which model is most supported by the data. For example if $\pi(y | \mathcal{M}_1)/\pi(y | \mathcal{M}_2) = 2$ that means that the probability of $y$ under model $\mathcal{M}_1$ is twice bigger than under model $\mathcal{M}_2$.  Jeffreys provided guidelines for qualitative interpretations of Bayes Factors \cite{kass1995bayes}, according to which, a BF of 4 and above is substantial evidence that $\mathcal{M}_1$ is the optimal model amongst the two under the Bayesian framework. 

We focus on the task of model choice within a range of factor models that span the spectrum from CFA to EFA which we describe in detail now. As described in Section \ref{sbfa_smc_sec:continuous_model}, researchers routinely express a factor hypothesis by setting certain loading parameters to zero, while freeing the rest. Such `exact zero' models, however, are very sensitive to model misspecifications and typically demonstrate poor fit in the presence of even small cross loadings, which are often found in real world data sets. For this reason researchers have questioned the suitability of such models, denoted by EZ henceforth, in research hypothesis testing \shortcite{SMSD15, HV18}. A proposed solution to this issue is to replace exact zero parameters with approximate zero (AZ) parameters through the use of prior distributions that are highly concentrated around zero. Such a model can be seen as the middle road between confirmatory and exploratory analysis. Specifically, in the classical confirmatory EZ models the loading matrix parameters are either free or zero. Freeing the zero parameters completely would result in an EFA model, so AZ offers a middle path where we free them while strongly constraining them to be near zero using appropriate prior distributions. The loading structures of all three models, exact zeros (EZ), approximate zeros (AZ), and exploratory factor model (EFA), are presented in Table \ref{sbfa_smc_table:model_loadings}. Note that the the first loading of each factor in EZ and AZ are set to $1$ for identifiability reasons, as the scale of the latent variable is non-identifiable. Similarly, exact zero models impose a diagonal structure in the residual covariance matrix, whereas the AZ model allows off-diagonal elements to be free but constrained near zero. For more information we refer the interested reader to \citeA{muthen2012} and \citeA{vamvourellis2021generalised}.

In the discussion above we have assumed that the true number of factors is known, since we generated the data ourselves using two factors. Hence, all three models so far (EZ, AZ and EFA) have been assumed to have two factors. In the absence of specific knowledge however, choosing the right number of factors remains an important challenge of applied factor analysis. To address this, we propose fitting EFA models with different number of factors and comparing their Bayes Factors. The most supported EFA model amongst them will likely be the one with the appropriate number of factors. That of course assumes a case of perfect model specification, as in Scenario 1. If there are cross loadings or correlated residual errors, such as in Scenario 2, it is possible that an EFA model with an extra factor would perform better because such an extra factor allows the model to accommodate the misspecifications. We demonstrate these cases by including the 1 factor EFA (EFA1) and the three factor EFA (EFA3) along with the two factor EFA (EFA2) and the rest of the models.

\subsubsection{Simulation Experiments}

Our starting point is the case study presented in Section \ref{sbfa_smc_subsec:contsim} using data generated according to Continuous Scenario 1. We fit the exact zero model (EZ) which we deliberately chose to match exactly the structure of the model that we used to generate the data. Hence we expect that EZ is supported by this data more than any other model, and indeed we find that it is. In the next scenario, we examine how model performance metrics change in the presence of small model misspecifications, a process we denote by `Continuous Scenario 2'. In particular, we generated data from the same model as before, except the loading structure introduces small cross loadings in three positions, as shown in Table \ref{sbfa_smc_table:loadings2}. 

\begin{table}[!htbp]
\centering
\begin{tabular}{*4c}
\toprule
 \multicolumn{2}{c}{Scenario 1} & \multicolumn{2}{c}{Scenario 2} \\
\midrule
$\Lambda_{:1}$ & $\Lambda_{:2}$ & $\Lambda_{:1}$ & $\Lambda_{:2}$ \\
\midrule
$1$     & $0$   & $1$   & $0$   \\
$.8$    & $0$   & $.8$  & $.3$  \\
$.8$    & $0$   & $.8$  & $0$   \\
$0$     & $1$   & $0$   & $1$   \\
$0$     & $.8$  & $.3$  & $.8$   \\
$0$     & $.8$  & $.3$  & $.8$   \\
\bottomrule
\vspace{.1cm} \end{tabular}
\caption{True factor loadings used in continuous data simulation.}
\label{sbfa_smc_table:loadings2}
\end{table}

\begin{table}[!htbp]
\centering
\begin{tabular}{*8c}
\toprule
 \multicolumn{2}{c}{EZ} & \multicolumn{2}{c}{AZ} & \multicolumn{2}{c}{EFA} \\
\midrule
$\Lambda_{:1}$ & $\Lambda_{:2}$ & $\Lambda_{:1}$ & $\Lambda_{:2}$ & $\Lambda_{:1}$ & $\Lambda_{:2}$ \\
\midrule
 $1$    & $0$   & $1$       & $\sim 0$ & $x$ & $x$ \\
 $x$    & $0$   & $x$       & $\sim 0$ & $x$ & $x$ \\
 $x$    & $0$   & $x$       & $\sim 0$ & $x$ & $x$ \\
$0$     & $1$   & $\sim 0$  & $1$      & $x$ & $x$ \\
$0$     & $x$   & $\sim 0$  & $x$      & $x$ & $x$ \\
$0$     & $x$   & $\sim 0$  & $x$      & $x$ & $x$ \\
\bottomrule
\vspace{.1cm} \end{tabular}
\caption{Loading structure of all three models, $x$ represents a free parameter, $\sim 0$ represents a free parameter with a prior distribution concentrated around $0$.}
\label{sbfa_smc_table:model_loadings}
\end{table}

Now we demonstrate the use of Bayes Factors for finding the optimal model amongst the ones considered under the Bayesian framework in the two scenarios discussed above. In Scenario 1 the EZ model matches exactly the data generation process so it is expected to be the best model. In Scenario 2, however, the loading structure imposed by the EZ does not exactly match the data generation process anymore. We expect that a more flexible factor model, such as the EFA, would perform better. Such flexibility comes at the cost of higher risk of overfitting, which occurs when a model fits noise rather than systematic patterns in the data. Furthermore, EFA models generally have slightly more free parameters than CFA models, and thus may result in lower estimation accuracy compared to EZ model. Hence, it is not clear a-priori whether the EFA model is better or worse than the EZ model in the presence of cross loadings in the data. The middle grown candidate model is the AZ, which preserves some of the structure of the EZ model but with some added flexibility. In the presence of small cross loadings we expect that the AZ model will be better than the EZ, but it is not clear how it would fair against the EFA.

Below we summarise the results of Scenario 1 in Tables \ref{sbfa_smc_table:results_sim1} and \ref{sbfa_smc_table:results_sim2} as it stands at the end of the inferential process (at point $i=200$). Table \ref{sbfa_smc_table:results_sim1} shows the log model evidence for each model from which we can deduce that EZ is the optimal model amongst those considered, since it has the highest value. In table \ref{sbfa_smc_table:results_sim2} we provide all the log Bayes Factors for each pair of models, which we will denote by $LBF$. For example the first column shows the log Bayes Factors $LBF(\text{EZ}/M) \coloneqq \log BF(\text{EZ}/M) = \log (\pi(y |\text{EZ}) / \pi(y |M))$ for $M \in \{ \text{AZ, EFA1, EFA2, EFA3} \}$. We can verify that all ratios are above $0$ which is a different way of saying that EZ is supported by the data the most. The second best model is AZ, as in the second column we can see it has positive ratios with the all other models but EZ. 

The table also shows us how we could have concluded that the right number of factors is two, in the absence of prior knowledge. We are looking for the EFA model that is most supported by the data amongst the three options. From Table \ref{sbfa_smc_table:results_sim1} we can see that EFA2 has the highest value of the three EFA models. Alternatively, we can verify this from Table \ref{sbfa_smc_table:results_sim2};  $LBF(\text{EFA1}/\text{EFA2}) = -58.2<0$ which means that EFA2 beats EFA1; and also $LBF(\text{EFA2}/\text{EFA3}) = 6.6>0$ which means that EFA2 also beats EFA3. 

\begin{table}[!htbp]
\centering
\begin{tabular}{*6c}
\toprule
Name & Log(Model Evidence) \\
\midrule
EZ   & ${\bf -1330.98}$  \\
AZ   & $-1331.27$   \\
EFA1 & $-1391.82$   \\
EFA2 & $-1333.67$  \\
EFA3 & $-1340.21$   \\
\bottomrule
\vspace{.1cm} \end{tabular}
\caption{Log Marginal Likelihood or Log Model Evidence for candidate models in simulation Scenario 1, at the final point $i=200$. We highlight model EZ with the highest value.}
\label{sbfa_smc_table:results_sim1}
\end{table}
\begin{table}[!htbp]
\centering
\begin{tabular}{*6c}
\toprule
{} & EZ & AZ & EFA1 & EFA2 & EFA3 \\
\midrule
EZ   & {}       & {}    & {}        & {}    \\
AZ   & $0.3$    & {}    & {}        & {}    \\
EFA1 & $60.8$   & $60.6$& {}        & {}    \\
EFA2 & $2.7$    & $2.4$ & $-58.2$   & {}    \\
EFA3 & $9.2$    & $8.9$ & $-51.6$   & $6.6$   \\
\bottomrule
\vspace{.1cm} \end{tabular}
\caption{Log Bayes Factor for candidate models in simulation Scenario 1, at the final point $i=200$. The table values represent the log ratio of the model on the top row divided by the model on the column. For example the $LBF(\text{EZ}/\text{AZ}) = 0.3$.}
\label{sbfa_smc_table:results_sim2}
\end{table}

Moving on, we now turn to the results in Scenario 2, where the data present cross-loadings. The results for Scenario 2 are summarised in Tables \ref{sbfa_smc_table:results_sim3} and \ref{sbfa_smc_table:results_sim4} where we can verify that the EZ model is no longer the optimal of the candidate models. In fact, the most supported model is now the AZ model since it has the highest marginal likelihood value in Table \ref{sbfa_smc_table:results_sim3}. This is a confirmation that the Bayes Factors are picking up on the fact that the exact zero model is no longer the best match for the data generation process, given the presence of the small cross loadings. It is interesting to gain a qualitative insight into how much does the data support each of the three models EZ, AZ and EFA2.    
If we exponentiate the values presented on Table \ref{sbfa_smc_table:results_sim4} we can read all four comparisons involving AZ from the table as follows: 

\begin{align*}
BF(\text{AZ})/BF(\text{EZ}) &= \frac{1}{\exp (LBF(\text{EZ})/\text{AZ}))} = \frac{1}{\exp (-9.8)} = 1\mathrm{e}{5}  \\
BF(\text{AZ}/\text{EFA2}) &=  \exp(2.6)  = 13.4
\end{align*}

We can verify that the hypothesised loading structure represented by EZ is too restrictive as AZ is clearly more supported. Additionally, because the cross loadings used to generate the data are relatively small, the loading structure is for the most part correct and for this reason the AZ model is strongly preferred also over the exploratory model with the correct number of factors EFA2.

Furthermore, we can note, that even in the presence of cross loadings we can deduce the right number of factors by observing which EFA performs best in terms of Bayes Factors. As in the previous scenario, EFA2 beats both EFA1 and EFA3, hence the evidence is pointing decisively towards EFA2 and two factors. Interestingly, EFA3 is also supported to a degree as the Bayes Factor $BF(\text{EFA2}/\text{EFA3})=  \exp(1.2)=3.3$ is relatively low at the value of $3$. That is probably happening because the additional factor is able to accommodate some of the cross loadings. 

\begin{table}[!htbp]
\centering
\begin{tabular}{*6c}
\toprule
Name & Log(Model Evidence) \\
\midrule
EZ   &  $-1330.69$  \\
AZ   & ${\bf -1290.85}$  \\
EFA1 & $-1347.83$   \\
EFA2 & $-1293.43$   \\
EFA3 & $-1294.61$   \\
\bottomrule
\vspace{.1cm} \end{tabular}
\caption{Log Marginal Likelihood or Log Model Evidence for candidate models in simulation Scenario 2, at the final point $i=200$. We highlight model AZ
with the highest value.}
\label{sbfa_smc_table:results_sim3}
\end{table}
\begin{table}[!htbp]
\centering
\begin{tabular}{*6c}
\toprule
{} & EZ & AZ & EFA1 & EFA2 & EFA3 \\
\midrule
EZ   & {}       & {}    & {}        & {}    \\
AZ   & $-9.8$   & {}    & {}        & {}    \\
EFA1 & $47.1$   & $57.0$& {}        & {}    \\
EFA2 & $-7.3$   & $2.6$ & $-54.4$   & {}    \\
EFA3 & $-6.1$   & $3.8$ & $-53.2$   & $1.2$   \\
\bottomrule
\vspace{.1cm} \end{tabular}
\caption{Log Bayes Factor for candidate models in simulation Scenario 2, at the final point $i=200$. The table values represent the log ratio of the model on the top row divided by the model on the column. For example the $LBF(\text{EZ}/\text{AZ}) = -9.8$.}
\label{sbfa_smc_table:results_sim4}
\end{table}

So far we have focused on model evaluation after having fit the model to the full batch of the data. However, the sequential framework we propose allows us to dynamically evaluate the models at each data point if we wanted. In practice it is not uncommon for sequential algorithms to be slow to reach a good set of particles if initiated at the first data point. One solution is to use the Adaptive Tempering algorithm which is designed to mitigate this issue exactly, see for example \cite{KBJ14,SC13}. Another solution is to initiate the sequential process particles using the output of a batch MCMC run using the first few data points. In this work, for demonstration purposes, we chose to initialise using the first 30 data points for all our runs. In general, further work is required to choose the size of the batch used for the initialisation step. After the initialisation step, the particles achieve good values (values of high likelihood) and remain stable from then on. The inference output suffices to carry out the comparisons between the models we described above at each new data point we process.

\begin{figure}
\centering
\includegraphics[width=0.8\textwidth]{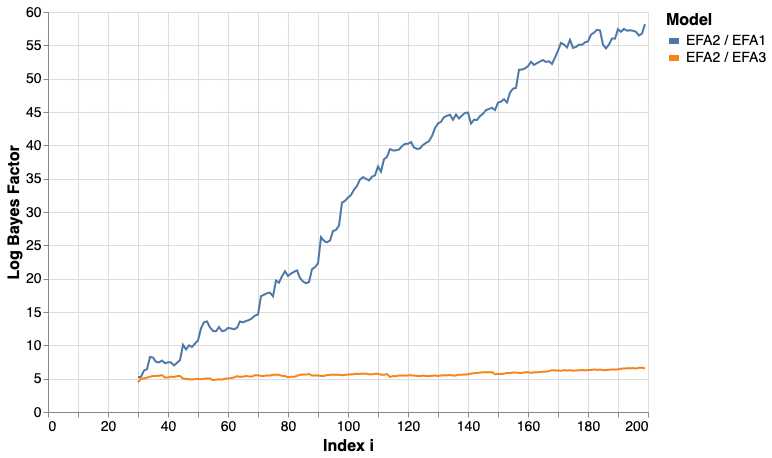}
\caption{Bayes Factors for EFA models of 1, 2 and 3 factors respectively in Scenario 1. We see that EFA2 outperforms both of the other models, which is an indication that the right number of factors is 2.}
\label{sbfa_smc_fig:ezsim_numberoffactors}
\end{figure}

\begin{figure}
\centering
\includegraphics[width=0.8\textwidth]{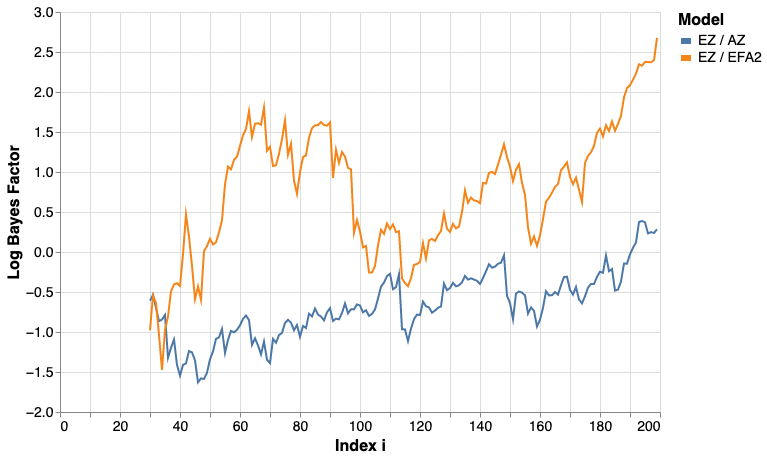}
\caption{Bayes Factor values of 3 candidate models, in Scenario 1 where the data was generated from a structure with zero cross loadings. We expect EZ to be the model of choice since it matches the structure of the data generation model the best of all the candidate models.}
\label{sbfa_smc_fig:ezsim_bf}
\end{figure}

We first address the question of how to choose the right number of factors to choose in Scenario 1. Recall that we initiate the sequential paradigm after having initialised the particles by running a batch MCMC on the first 30 points. We present the chart of log Bayes Factors between the EFA models in Figure \ref{sbfa_smc_fig:ezsim_numberoffactors} where the horizontal axis shows the index of the data point and the vertical axis shows the Log Bayes Factor. We can confidently claim that the right number of factors is two based on the fact that EFA2 outperforms easily both EFA1 and EFA3. We present the Sequential Log Bayes Factors amongst the rest of the models for Scenario 1 in Figure \ref{sbfa_smc_fig:ezsim_bf}. We choose to present all the factors with EZ in the numerator and the other 2 models in the denominator. The chart highlights two interesting points. First, it shows that the AZ and EZ models present very comparable level of support throughout the dataset as their log factor ratio stays near zero throughout. Second, we see that the EZ model starts to outperform the best EFA model, EFA2, after about 50 points when their log ratio turns positive. However, it does not surpass the notional mark of $1.4$ until it hits point 180. We note that  $BF(\text{EZ}/\text{EFA2})=\exp(1.4)=4$ and recall that a Bayes Factor of $4$ is considered substantial support in favour of the better model.

\begin{figure}
\centering
\includegraphics[width=0.8\textwidth]{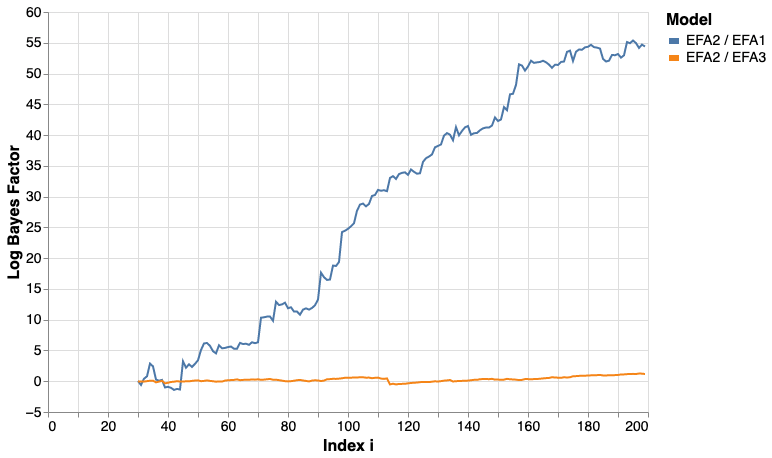}
\caption{Bayes Factors for EFA models of 1, 2 and 3 factors respectively in Scenario 2. We see that EFA2 outperforms both of the other models, which is an indication that the right number of factors is 2. However, because of model misspecification the EFA model with an additional factor, EFA3, remains competitive.}
\label{sbfa_smc_fig:clsim_numberoffactors}
\end{figure}

\begin{figure}
\centering
\includegraphics[width=0.8\textwidth]{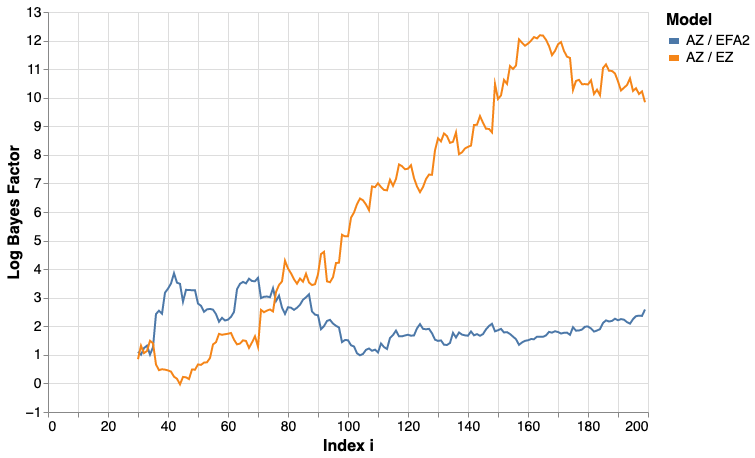}
\caption{Bayes Factor values of the 4 candidate models, in Scenario 2 where the data was generated from a structure that included cross loadings. We expect AZ to be the model of choice since it matches the structure of the data generation model the best of all the candidate models.}
\label{sbfa_smc_fig:clsim_bf}
\end{figure}

In Figure \ref{sbfa_smc_fig:clsim_numberoffactors} we present the Sequential Log Bayes Factors for the EFA models in Scenario 2 in order to choose the right number of factors. We see that EFA2  outperforms EFA1 from the beginning but $LBF(\text{EFA2}/\text{EFA3})$ remains above $0$ but below $2.5$ for the rest of the dataset. This would be a sign that the right number of factors is probably $2$, but there are additional model misspecifications that might call for an additional factor. More generally, when the data support an EFA model with $K$ factors only slightly more an EFA with $K+1$ factors, it could be an indication that the true number of factors is $K$ but there are model misspecifcations that justify using the EFA model with $K+1$ factors as well. We then present the comparisons of the rest of the models in Figure \ref{sbfa_smc_fig:clsim_bf} where we choose to present all the ratios with AZ in the numerator. The EZ model remains competitive for approximately the first 100 points, and after that the log ratio grows decisively above $5$. The ratios with EFA2 favours AZ in the range of $2-3$ throughout which confirms the interpretation we discussed in relation to Table \ref{sbfa_smc_table:results_sim4}. 

\subsection{Application: Big 5 Personality Test - British Household Panel Survey}\label{sbfa_smc_subsec:big5}

In this Section we apply the proposed framework to a real word dataset to highlight the benefits of sequential model selection in the case where the true data generation process is unknown. The data set we work with comes from the British Household Panel Survey in 2005-06, which concentrated on female subjects between the ages of 50 and 55; the sample size consists of 676 individuals.  The `Big 5 Personality Test', as it is known, is a $15$-item questionnaire on topics of social behaviour and emotional state. Each item receives an answer from each participant on a scale from $1-7$, $1$ being `strongly disagree' and $7$ being `strongly agree'. The test is meant to measure five major, potentially correlated, personality traits. Each factor corresponds to one trait, and is hypothesised to explain exactly 3 out of 15 items. Typically, as is the case in our analysis, the questions are ordered in a way such that the first three questions correspond to the first factor, the next three questions correspond to the second factor; the pattern continues so that the last three questions correspond to the fifth factor. For our analysis we standardised the data by removing the mean and set the standard deviation to $1$ so that we receive standardised loading values. 

Prior research has suggested that the dataset demonstrates potentially small cross loadings, as well as correlated residual errors, possibly as a result of negative wordings of some of the items in the questionnaire. As a result, the exact zero model has been found to have poor fit, while the equivalent AZ model fits the data much better \shortcite{muthen2012, vamvourellis2021generalised}. We are also interested in examining whether more flexible models are more supported by the data, to benchmark against the confirmatory models with the hypothesised structure. Since the questionnaire was constructed to measure 5 personality traits, we will include the 5 factor EFA model which will perform better than the rest if the hypothesised factor structure is not correct. Finally, we include the most flexible model possible, the saturated model denoted by `SAT', that imposes no factor structure on the dataset. 

\begin{figure}
\centering
\includegraphics[width=0.8\textwidth]{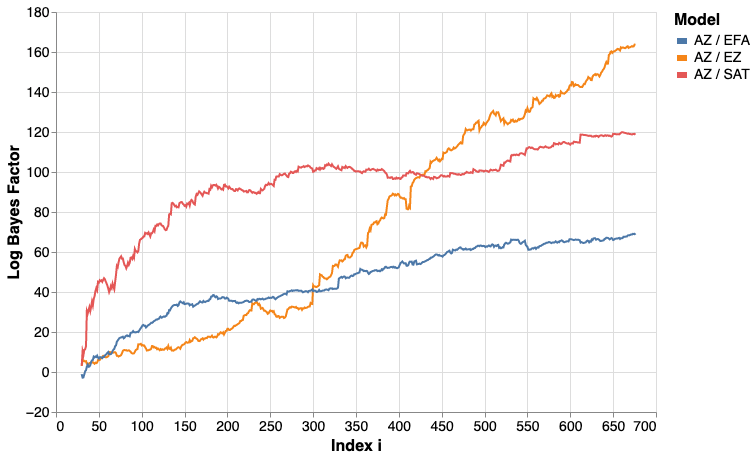}
\caption{Bayes Factor values of the 4 candidate models. Based on prior research we expect AZ to be the model of choice.}
\label{sbfa_smc_fig:muthen_bf}
\end{figure}

In the sequential model comparison chart, Figure \ref{sbfa_smc_fig:muthen_bf}, it becomes clear early on that the hypothesised factor structure is correct. Even though the cross loadings and residual errors cause the EZ model to not perform optimally in terms of Bayes Factors, the overall loading structure is supported by the data. We can conclude this because the most supported model is the AZ model, which easily outperforms EFA and SAT, both of which are presumed to be more flexible. Their flexibility would result in higher model evidence if indeed the factor structure hypothesised in AZ was wrong. The fact that the most flexible models fail to perform better is an indication that the hypothesised structure of AZ is supported. The SBF framework can facilitate dynamic adjust to surveys, such as the one analysed here. As we can see,it becomes immediately clear that the AZ is better than the EZ, say after the first 80 participants. If the analysis was done in real time, the researchers could have stopped the study early and potentially revised the structure or the wording of the questionnaire. 

\begin{figure}
\centering
\includegraphics[width=1\textwidth]{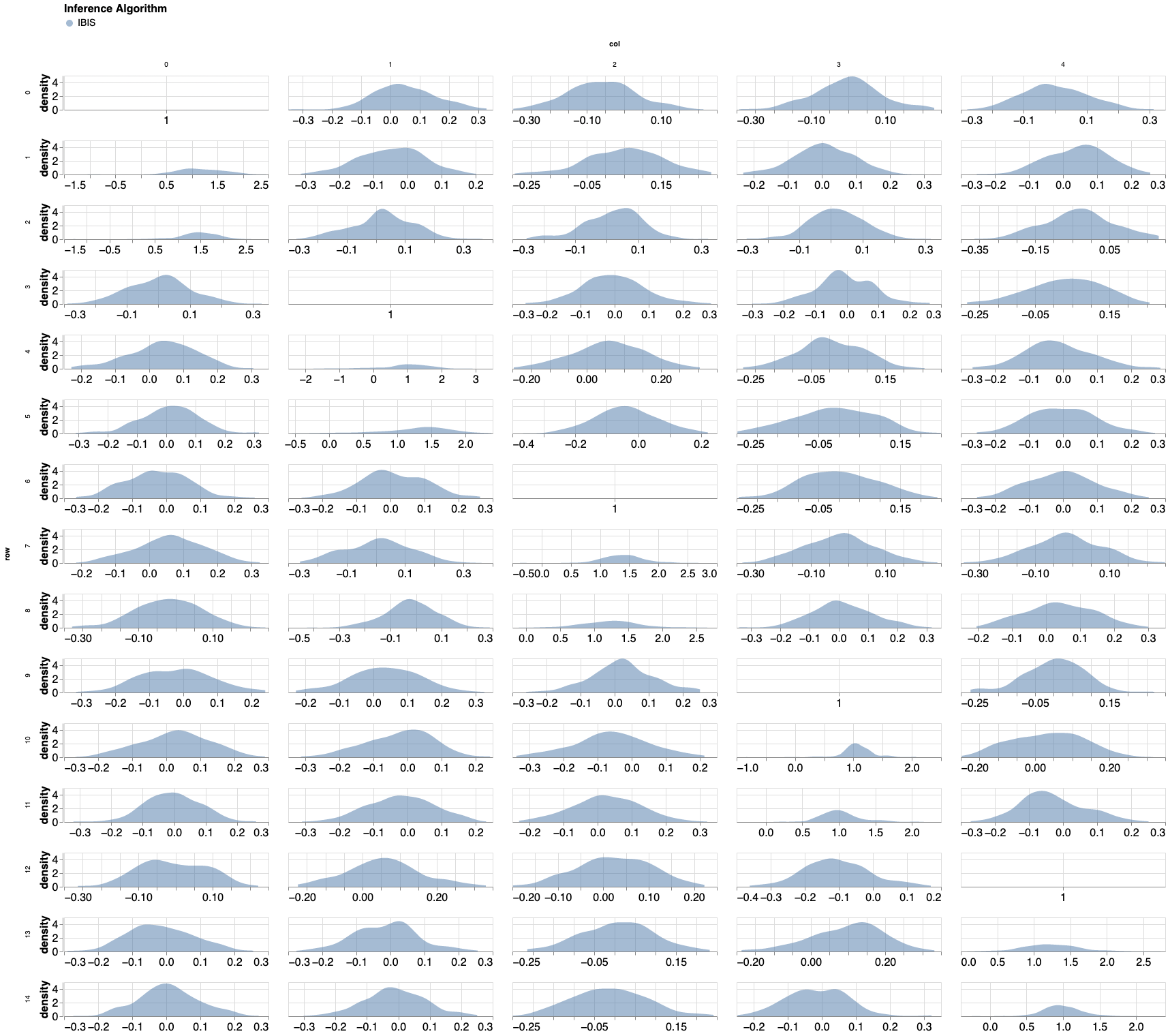}
\caption{Posterior density plots for the loadings parameter of the AZ model fit to the Big5 dataset.}
\label{sbfa_smc_fig:muthen_loadings}
\end{figure}

Having selected the right model we can also draw posterior samples at each point and learn the loadings of the factors for each item. Here we present the posterior mean estimated loadings after the last data point has been processed in Table \ref{sbfa_smc_table:big5_loadings}. We can see that some cross loadings, such as 1-st factor 8-th item are estimated to be non-zero, despite the strong priors towards zero. These items are prime candidates for sources of model misspecification. Overall, all such cross loadings are estimated to be smaller than $0.1$ in absolute value while the main loading structure takes values around $1$. A posterior density plot of the draws, Figure \ref{sbfa_smc_fig:muthen_loadings} can reveal some further characteristics of the areas of misfit. For example, some cross loading densities, such as the the 4-th factor on the 14-th item, present substantial skewness to the right, which means that there is potentially a stronger source of cross loading than the cross loading of the 1-st factor on the 8-th item which is a more symmetric density plot. 

\begin{table}[!htbp]
\centering
\begin{tabular}{*5c}
\toprule
$\Lambda_{:1}$ & $\Lambda_{:2}$ & $\Lambda_{:3}$ & $\Lambda_{:4}$ & $\Lambda_{:5}$ \\
\midrule
    1.& 0. & -0.1 & -0.& 0.   \\ 
    1.2 &-0.& 0.1 & 0.& 0.   \\ 
    1.5 & 0.& 0.& 0. & -0.   \\ 
    0.& 1. & -0. & -0.& 0.   \\ 
    0.& 0.9 & 0.1 & 0. & -0.   \\ 
    0.& 1.3 & -0. & -0. & -0.   \\ 
   -0.& 0.& 1. & -0. & -0.   \\ 
    0.1 & 0.& 1.4 & -0. & -0.   \\ 
   -0. & -0.& 1.3 & 0.& 0.   \\ 
   -0.& 0.& 0.& 1.& 0.   \\ 
    0. & -0. & -0.1 & 1.1 & 0.   \\ 
   -0. & -0.& 0.& 1. & -0.   \\ 
   -0.& 0.1 & 0. & -0.1 & 1.   \\ 
   -0. & -0.& 0.& 0.1 & 1.3  \\ 
    0. & -0. & -0. & -0.& 1.   \\ 
\bottomrule
\vspace{.1cm} \end{tabular}
\caption{Posterior mean estimates of loading values for the Big 5 dataset at the final point of inference. We present the values at 1 decimal point.}
\label{sbfa_smc_table:big5_loadings}
\end{table}

\section{Discussion}\label{sbfa_sec:discussion}

In this paper we propose an efficient sequential scheme for factor analysis with continuous latent variables. The modelling case of continuous Normally distributed data can be handled using the existing framework of IBIS \cite{Chopin2002a}. The crucial fact is that it is possible to write down the distribution of the data, marginalising out the latent variables. However, when that marginal distribution is not available, as is the case in IRT models, the standard IBIS algorithm does not apply. We develop an efficient scheme based on IBIS that can handle categorical and non-normally distributed continuous data, using the Laplace and Variational Bayes approximations. Other SMC$^2$ based approaches, while wholly valid, they tend to be inefficient in the context of factor analysis. The reason is that they are constructed to work with latent variables that present temporal correlations, whereas factor modelling usually involves independent latent variables. Our scheme takes advantage of the independence structure to offer an efficient alternative. Merging the fields of Sequential Monte Carlo methods and Factor Analysis could be a fertile area of research ahead. Additionally, future practitioners stand to benefit from more research into the implementation details of these frameworks. Moreover, we note that the sequential inferential framework proposed in this work provides a robust computational alternative to MCMC schemes for factor analysis even in the non-sequential setting. An interesting future research direction is a more formal performance comparison between the two approaches. 

Finally, another area of future research is that of sequential model choice. The sequential framework proposed here facilitates model choice in two ways. While in this work we focus on Bayes Factors, scoring rules is a powerful alternative tool. Scoring rules assess a model via predictive performance, even when when prediction is not amongst the research objectives. The philosophical basis of this approach suggests that a model fits the data well when it is able to predict new data that was not used for estimating the model parameters. Scoring rules evaluated on out-of-sample data are available as a direct output of the data tempering strategy adopted in this work, under the prequential framework \cite{DM14}. A formal comparison of Sequential Bayes Factors and such scoring rules is an interesting field of further research.

\vspace{\fill}

\bibliographystyle{chicago}
\bibliography{references}

\appendices
\section{HMC Implementation Details}\label{sbfa_app:code}

One can also improve efficiency by providing good values Hamiltonian MCMC tuning parameters. In particular, the mass matrix needed to propose new values can be estimated accurately by running a longer chain for the first particle, which can then be reused as an initial value for the rest the particles' chains. Similarly, any other MCMC parameters that require a long adaptation phase, such as the step size in HMC, can be learned by running one long chain for the first particle, then used as initial values for the rest of the particles. With this approach, each time the criterion indicates a resample, we run one a long chain for the first particle and then we can afford to run short chains for the rest of the particles. In our experiments running the first particle for 500 steps produced sufficiently accurate initial values for the rest to be run for 10 steps. Further efficiencies can be achieved by running the particle chains in parallel as they are independent of each other.

\section{IBIS with Laplace Approximation}\label{sbfa_smc_appendix:laplace}
In this section we derive the approximation to posterior $\pi(z_i|y_i, \theta)$ via the Laplace method. The goal is to use the approximating distribution, $p^L(z_i|y_i, \theta)$ in place of the proposal $q(.)$ Our model assigns $N(0,4)$ priors to each component of $z$ and also assumes a-priori independence between these components. More generally we can note, even though we do not need it in this derivation, that based on this prior and \eqref{sbfa_smc_bin_augmented}, conditional on $\theta$, the $z_i$ rows are independent a-posteriori. Thus, in order to approximate conditional posterior of $z$ given $\theta$, one can approximate the corresponding posteriors of each $z_{i\cdot}$ separately. For the exposition that follows we can drop the subscripts, since the focus of the approximation is the $i$-th point $z_i$ only, so we adopt the running assumption that $z$ and $y$ represent $z_i$ and $y_i$ for the remaining of the section. We focus on approximating the posterior 
\begin{equation}
\label{sbfa_smc_eq:IndivPost}
\pi(z\, | \,y,\theta) \propto f(y \, | \, z,\theta)\exp(-\tfrac{1}{2}zz^{T}).
 \end{equation}
We then target the logarithm of \eqref{sbfa_smc_eq:IndivPost}, 
\begin{equation}
\label{sbfa_smc_eq:logtarget}
\ell(z\, | \,y,\theta) = \log f(y\, | \,z, \theta) -\tfrac{1}{2} zz^{T}= \sum_{j=1}^p \left[y_{j}\log \pi_{j}(z,\theta) + (1-y_j)\log\left\{1-\pi_{j}(z,\theta)\right\}\right] -\tfrac{1}{2}\sum_{\ell=1}^k z_{\ell}^2
\end{equation}
In order to apply the Laplace approximation on \eqref{sbfa_smc_eq:IndivPost} we need the first and second derivatives of \eqref{sbfa_smc_eq:logtarget} with respect each $z_{\ell}$ for $\ell=1,\dots,k$. These are 
\begin{eqnarray}
\tfrac{\partial}{\partial z_{\ell}} \ell(z\, | \,y,\theta) &=&  - z_{\ell}+\sum_{j=1}^p \left\{\frac{y_j \tfrac{\partial}{\partial z_{\ell}}\pi_{j}(z,\theta)}{\pi_{j}(z,\theta)} - \frac{(1-y_j)\tfrac{\partial}{\partial z_{\ell}}\pi_{j}(z,\theta)}{1-\pi_{j}(z,\theta)}\right\}  \nonumber\\
&=& - z_{\ell}+ \sum_{j=1}^p \tfrac{\partial}{\partial z_{\ell}}\pi_{j}(z,\theta) \left\{ \frac{y_j }{\pi_{j}(z,\theta)} - \frac{1-y_j}{1-\pi_{j}(z,\theta)}\right\}  
\label{sbfa_smc_eq:score}
\end{eqnarray}
\begin{eqnarray*}
\tfrac{\partial^2}{\partial z_{\ell}^2} \ell(z\, | \,y,\theta) &=&  -1 +\sum_{j=1}^p \tfrac{\partial^2}{\partial z_{\ell}^2}\pi_{j}(z,\theta) \left\{ \frac{y_j }{\pi_{j}(z,\theta)} - \frac{1-y_j}{1-\pi_{j}(z,\theta)}\right\}  \nonumber\\
&+& \sum_{j=1}^p\tfrac{\partial}{\partial z_{\ell}}\pi_{j}(z,\theta) \left[ -\frac{y_j \tfrac{\partial}{\partial z_{\ell}}\pi_{j}(z,\theta)}{\pi_{j}(z,\theta)^2} - \frac{(1-y_j)\tfrac{\partial}{\partial z_{\ell}}\pi_{j}(z,\theta)}{\left\{1-\pi_{j}(z,\theta)\right\}^2}\right]  \nonumber\\
&=&-1 +\sum_{j=1}^p \tfrac{\partial^2}{\partial z_{\ell}^2}\pi_{j}(z,\theta) \left\{ \frac{y_j }{\pi_{j}(z,\theta)} - \frac{1-y_j}{1-\pi_{j}(z,\theta)}\right\} \nonumber\\
&-& \sum_{j=1}^p\left\{\tfrac{\partial}{\partial z_{\ell}}\pi_{j}(z,\theta)\right\}^2 \left[ \frac{y_j }{\pi_{j}(z,\theta)^2} + \frac{1-y_j}{\left\{1-\pi_{j}(z,\theta)\right\}^2}\right]
\end{eqnarray*} 
and for $m=1,\dots,k,$ with $\ell\neq m$
\begin{eqnarray*}
\tfrac{\partial^2}{\partial z_{\ell}\partial z_m} \ell(z\, | \,y,\theta) &=&  \sum_{j=1}^p \tfrac{\partial^2}{\partial z_{\ell}\partial z_m}\pi_{j}(z,\theta) \left\{ \frac{y_j }{\pi_{j}(z,\theta)} - \frac{1-y_j}{1-\pi_{j}(z,\theta)}\right\}  \nonumber\\
&-& \sum_{j=1}^p\tfrac{\partial}{\partial z_{\ell}}\pi_{j}(z,\theta) \tfrac{\partial}{\partial z_{m}}\pi_{j}(z,\theta)\left[ \frac{y_j }{\pi_{j}(z,\theta)^2} + \frac{1-y_j}{\left\{1-\pi_{j}(z,\theta)\right\}^2}\right] 
\end{eqnarray*} 
The above can also provide the Fisher's information matrix $\mathcal{I}(z\, | \, \theta)$, as for $\ell=1,\dots,k$, we get
\begin{eqnarray}
\left[\mathcal{I}\right]_{\ell\ell}(z\, | \, \theta) &=& - E\left\{\tfrac{\partial^2}{\partial z_{\ell}^2} \ell(z\, | \,y,\theta)\right\} = 1 + \sum_{j=1}^p\left\{\tfrac{\partial}{\partial z_{\ell}}\pi_{j}(z,\theta)\right\}^2 \left\{ \frac{1 }{\pi_{j}(z,\theta)} + \frac{1}{1-\pi_{j}(z,\theta)}\right\} \nonumber \\
&=& 1+\sum_{j=1}^p \frac{\left\{\tfrac{\partial}{\partial z_{\ell}}\pi_{j}(z,\theta)\right\}^2 }{\pi_{j}(z,\theta)\left\{1-\pi_{j}(z,\theta)\right\}}
\label{sbfa_smc_eq:Iii}
\end{eqnarray}
and for $m=1,\dots,k,$ with $\ell\neq m$
\begin{eqnarray}
\left[\mathcal{I}\right]_{\ell m} (z\, | \, \theta)&=& - E\left\{\tfrac{\partial^2}{\partial z_{\ell}\partial z_m} \ell(z\, | \,y,\theta)\right\} = \sum_{j=1}^p \frac{\tfrac{\partial}{\partial z_{\ell}}\pi_{j}(z,\theta) \tfrac{\partial}{\partial z_{m}}\pi_{j}(z,\theta)}{\pi_{j}(z,\theta)\left\{1-\pi_{j}(z,\theta)\right\}}
\label{sbfa_smc_eq:Iij}
\end{eqnarray}
It remains to calculate $\pi_{j}(z,\theta)$ and $\tfrac{\partial}{\partial z_{\ell}}\pi_{j}(z,\theta)$. Based on the model in $\eqref{sbfa_smc_bin_augmented2}$ we get
\begin{eqnarray*}
 \pi_{j}(z,\theta) = \frac{\exp\left(\alpha_j + \sum_{\ell=1}^k z_{\ell} \Lambda_{\ell j}\right)}{1+\exp\left(\alpha_j + \sum_{\ell=1}^k z_{\ell} \Lambda_{\ell j}\right)}, &
 \tfrac{\partial}{\partial z_{\ell}}\pi_{j}(z,\theta) = \frac{\exp\left(\alpha_j + \sum_{\ell=1}^k z_{\ell} \Lambda_{\ell j}\right)\Lambda_{\ell j}}{\left\{1+\exp\left(\alpha_j + \sum_{\ell=1}^k z_{\ell} \Lambda_{\ell j}\right)\right\}^2},
\end{eqnarray*}
%\sum_{\ell=1}^k 
which we can plug in to \eqref{sbfa_smc_eq:score}, \eqref{sbfa_smc_eq:Iii} and \eqref{sbfa_smc_eq:Iij} to obtain the Laplace approximation
\begin{equation}
\label{sbfa_smc_eq:laplace-form}
z \; \sim\; N\left\{ \underset{z}{\arg\max} \;\ell(z\, | \,y,\theta),\;  \mathcal{I}(z\, | \, \theta)^{-1}\right\}
\end{equation}

where $\underset{z}{\arg\max} \;\ell(z\, | \,y,\theta)$ can be obtained via the Fisher's Scoring algorithm.

Alternatively we can compute the Observed Information matrix which is negative the Hessian of the log-likelihood evaluated at the mode. For that we will need the second derivative of $\tfrac{\partial^2}{\partial^2 z_{\ell}}\pi_{j}(z,\theta)$ as follows: 
\begin{eqnarray*}
\tfrac{\partial}{\partial z_{\ell}}\pi_{j}(z,\theta) = \frac{-\Lambda^2_{\ell j} \exp\left(\alpha_j + \sum_{\ell=1}^k z_{\ell} \Lambda_{\ell j}\right)}{\left\{1+\exp\left(\alpha_j + \sum_{\ell=1}^k z_{\ell} \Lambda_{\ell j}\right)\right\}^2}
\end{eqnarray*}
which we can plug into equation \eqref{sbfa_smc_eq:score} to compute the matrix. 

%\vfill\eject
\end{document}